\documentclass[10pt]{article}
\usepackage[margin=0.85in]{geometry}
\usepackage[T1]{fontenc}
\usepackage{lmodern}
\usepackage{graphicx}
\usepackage{float}
\usepackage{longtable}
\usepackage{array}
\usepackage{booktabs}
\usepackage{ragged2e}
\usepackage{xurl}
\usepackage{hyperref}
\usepackage{natbib}
\usepackage{xcolor}
\hypersetup{colorlinks=true, linkcolor=blue, citecolor=blue, urlcolor=blue}
\title{The Serialized Bridge: Understanding and Recovering LLM Serving Performance under Blackwell GPU Confidential Computing}
\author{Hang Yin\\\texttt{hangyin@phala.network} \and Kevin Wang\\\texttt{kvinwang@phala.network}}
\date{May 2026}
\begin{document}
\maketitle

\begin{abstract}

GPU Confidential Computing (GPU-CC) now preserves GPU-local performance: on
NVIDIA B300, BF16 matmul runs at 0.998x of non-confidential performance. Yet
LLM serving under Intel TDX plus GPU-CC still loses 13-27\% of throughput, and
KV-cache restore latency can more than double. This paper studies that gap on
two Blackwell platforms, RTX Pro 6000 and B300 HGX, and identifies its dominant
cause: the confidential VM-GPU bridge, not GPU compute.

We find that GPU-CC turns host/device movement into a serialized,
high-setup-cost channel. Secure copies do not gain CUDA-stream concurrency
within a context, asynchronous transfers block at the runtime boundary, and
small crossings pay a fixed toll. This violates the assumptions of modern
inference runtimes, where DMA is expected to be cheap, concurrent, and
asynchronous. In vLLM dense decode, the gap closes around 44x-slower small
alloc-and-copy operations; targeted patches reject alternative explanations.
A scheduling flag recovers 57\% of the gap, while a worker-thread drain recovers
up to 92\% in qualified high-concurrency runs. The same bridge model explains a
+131\% KV-restore penalty and a 34x model-load slowdown.

Blackwell also changes the confidential tenancy unit. We qualify confidential
multi-GPU NVSwitch tenants on B300, including 510 GB/s NVLink P2P inside a CVM
and concurrent isolated tenants, and identify the remaining fabric-attestation
gap for production confidential AI platforms.

\end{abstract}
\section{Introduction}

The question that launched GPU confidential computing benchmarking - "how much
slower is the GPU in confidential mode?" - now has a settled answer: barely.
Hopper-era studies found GPU-local LLM compute within a few percent of native
\citep{hopper-cc-benchmark,performance-cc-gpus}, and our Blackwell measurements agree to three
decimal places: BF16 matmul on B300 is 0.998x under CC, and a CUDA graph
chaining 96,000 matmuls back-to-back is 1.0012x. The GPU, its tensor cores,
and its HBM are not where confidential computing costs anything.

And yet the end-to-end numbers are not settled at all. On the same B300, dense
LLM decode loses 13-22\% under CC depending on configuration, mixture-of-experts
serving loses up to 25\%, KV-cache restore latency rises 131\%, and loading
GPT-OSS-120B takes 287 seconds with the default loader - on hardware that can
move the same bytes in well under a minute. Worse, the relative penalty is
growing across hardware generations: the same vLLM stack that loses about 10\%
on an RTX Pro 6000 loses about 26\% on the much faster B300, because everything
the GPU accelerates shrinks except the confidential data path. An H200
boundary experiment separates mechanism from consequence: the secure-copy floor is
already present on Hopper, but Blackwell is where the relative tax becomes
large enough to flip runtime policy.

This paper locates all of these losses in one place and explains them with one
mechanism:
\textbf{under GPU-CC, the bridge between the confidential VM and the GPU is serialized.}
Three measured properties define the behavior:

\begin{enumerate}
\item \textbf{No concurrency within a context.} All cross-device traffic in a CUDA
context shares a small fixed pool of secure copy channels. Sixteen CUDA
streams each issuing 32-byte device-to-host copies achieve approximately
zero parallelism under CC (40 to 39 microseconds per copy from one stream
to sixteen), versus 24\% scaling with CC off. Additional bandwidth is only
reachable by adding CUDA \emph{contexts}: one context sustains 0.2x of native
transfer bandwidth, while 24 contexts recover to 0.6-0.7x.
\item \textbf{Asynchrony is silently revoked.} \path|copy_(..., non_blocking=True)| to
pinned memory - the standard PyTorch idiom for overlapping transfers -
blocks the calling CPU thread for the full transfer under CC. The hint is
accepted by the API and neutralized below it.
\item \textbf{Every crossing pays a fixed toll.} TDX guest private memory cannot be
read by GPU DMA, so each host-to-device copy crosses an encrypted staging
path with a setup cost we measure at roughly 330 microseconds per call -
three orders of magnitude above the per-byte cost for small payloads.
\end{enumerate}

None of these properties slows a workload that computes in place. All of them
punish software that crosses the bridge often, in small pieces, or, most
damagingly, \emph{in parallel on purpose}. Modern inference runtimes do exactly
that, because
on non-confidential GPUs, overlap is free performance. The result is what we
call \textbf{policy inversion}: a scheduling optimization that helps without CC
hurts with CC. vLLM's default async scheduling saves about 3 ms per decode
step without CC by overlapping output drain with next-step preparation; under
CC the same policy \emph{costs} about 4 ms per step, because the overlapped copies
serialize anyway and the runtime pays stream-arbitration overhead for an
overlap that never happens.

Because the mechanism is software-visible, it is software-fixable. We show a
recovery hierarchy measured on B300: a one-line flag (\texttt{--no-async-scheduling})
closes 57\% of the dense-decode CC gap and essentially all of the KV-restore
gap; a 30-line patch that moves the blocking drain to a worker thread restores
true pipelining and closes up to 92\% of the gap at high concurrency; and a
CC-aware model loader that pools secure contexts cuts GPT-OSS-120B load time
from 287 s to 8.4 s. Each fix follows directly from the serialization model,
and each transfers across the two Blackwell platforms we measure.

Finally, Blackwell changes the \emph{capability} boundary as well as the
performance boundary. On B300 HGX, a confidential tenant is no longer one
passed-through device but a partition of an NVSwitch fabric. We report the
first public qualification of this mode: confidential 2-GPU tenants with full
NVLink P2P at 510 GB/s, concurrent isolated tenants on disjoint partitions,
a fixed 1/2/4/8 partition vocabulary that becomes the scheduling API, and a
trust gap - Fabric Manager identity and NVSwitch routing state remain outside
tenant attestation evidence.

\subsection{Contributions}

\begin{enumerate}
\item \textbf{A causal performance model of the GPU-CC bridge.} Targeted
microbenchmarks on two Blackwell platforms establish that secure copy
channels serialize per-context traffic, that asynchronous copies block, and
that per-crossing setup dominates small transfers - while compute and
GPU-local memory stay at parity. Crypto-throughput and HBM-cipher
explanations are ruled out by instruction-set ablations and chained-graph
experiments. A matched H200 experiment repeats the same bridge signatures,
showing that the bridge law is not a B300-only artifact.
\item \textbf{Root cause and recovery for confidential LLM serving.} A profiler
accounting loop attributes the vLLM dense-decode CC gap to 44x-slower
alloc-and-copy calls (1,138 calls explain 1.54 s of a 1.56 s slowdown);
single-variable patch experiments refute each alternative
explanation; and we demonstrate policy inversion of async scheduling on
Blackwell with a measured recovery hierarchy: 57\% from one flag, up to 92\%
from a worker-thread patch at c=512, with KV-churn TTFT recovering from
+131\% to +2\%. The matched H200 result confirms that CC removes async's
benefit, but there it stops at parity rather than becoming a harmful
default.
\item \textbf{CC-aware data-movement engineering that transfers across platforms.} A
secure-context-pooling loader with a measured lifecycle cost model (5.2 s
context creation, 3.9 s destruction per 8-worker load) cuts GPT-OSS-120B
weight load 34x, within 5\% of identical on both Blackwell platforms;
reuse-aware KV-offload policy cuts warm TTFT 2.97x under CC.
\item \textbf{First public qualification of B300 confidential fabric tenants}, with
measured NVLink P2P inside a CVM (510 GB/s), concurrent tenant isolation,
the partition vocabulary that constrains scheduling, and an analysis of
the remaining fabric-attestation gap.
\end{enumerate}

\section{Background}

\subsection{GPU-CC and the CVM-GPU Bridge}

In the configuration we study, Intel TDX protects a confidential VM
\citep{intel-tdx} and NVIDIA GPU-CC protects GPU execution and the GPU's Compute
Protected Region (CPR) \citep{nvidia-cc-architecture,nvidia-secure-ai}. We follow the
terminology of Gu et al.: CPU-CC protects CVM private memory, GPU-CC protects
the GPU domain, and the \emph{CVM-GPU bridge} is the runtime path between the two
protected domains \citep{gpu-cc-blueprint}.

The bridge is not ordinary DMA. The confidential CPU domain cannot expose its
private memory to the device, and the device cannot expose the CPR to the
host, so commands and data cross through shared staging (bounce) buffers with
encrypted or authenticated payloads \citep{nvidia-h100-cc-ga,gpu-cc-blueprint}. Two
consequences matter for performance. First, each crossing involves staging
setup and cipher work on the CPU side, a fixed per-call cost. Second, the
copies themselves are carried by \emph{secure copy channels}, a constrained
hardware resource. NVIDIA's operations guidance is explicit that the limit is
systemic: the maximum number of CUDA contexts under CC "arises from a
system-wide restriction on the number of secure copy channels that the GPU
DMA engines support" \citep{nvidia-secops}.

GPU memory inside the CPR is protected by access-control firewalling rather
than per-byte encryption \citep{gpu-cc-blueprint}, which is why GPU-local compute
and HBM traffic are nearly free under CC while bridge crossings are not. This
split - compute at parity, movement taxed - is the organizing fact of the
paper.

\subsection{Trust Model}

The tenant CVM relies on TDX for the CPU boundary and GPU-CC for the GPU
domain. The host remains responsible for launching CVMs, assigning devices,
and, on B300 HGX, operating the NVSwitch fabric control plane (Fabric Manager
and NVLSM) \citep{nvidia-fabric-manager,nvidia-shared-nvswitch}. The tenant can verify TDX evidence, GPU-CC
mode, GPU ready state, and guest-visible fabric health. It cannot yet bind the
Fabric Manager binary, its configuration, or the programmed NVSwitch routing
tables into attestation evidence. Section 7 measures what works under this
model and Section 7.3 analyzes the gap.

\subsection{From Hopper to Blackwell}

Hopper established commercial GPU-CC and its first performance lessons:
compute-heavy inference close to native, transfer-heavy paths taxed
\citep{nvidia-h100-cc-ga,hopper-cc-benchmark,performance-cc-gpus}. Hopper's multi-GPU answer was
Protected PCIe: assign a whole attached GPU complex to one tenant and treat
the fabric as private to that assignment. Blackwell B300 adds a different
mode: the host-managed NVSwitch fabric can expose \emph{partitions} - 1/2/4/8 GPU
sets - as confidential tenants while the fabric stays shared. Table 1
summarizes the three modes.

\begingroup
\scriptsize
\setlength{\tabcolsep}{2pt}
\setlength{\LTleft}{0pt}
\setlength{\LTright}{0pt}
\begin{longtable}{@{\extracolsep{\fill}}>{\RaggedRight\arraybackslash}p{0.158\linewidth}>{\RaggedRight\arraybackslash}p{0.174\linewidth}>{\RaggedRight\arraybackslash}p{0.268\linewidth}>{\RaggedRight\arraybackslash}p{0.300\linewidth}@{}}
\toprule
\textbf{Mode} & \textbf{Typical tenant shape} & \textbf{Fabric assumption} & \textbf{Scheduling implication} \\
\midrule
\endfirsthead
\toprule
\textbf{Mode} & \textbf{Typical tenant shape} & \textbf{Fabric assumption} & \textbf{Scheduling implication} \\
\midrule
\endhead
Single-GPU GPU-CC & 1 GPU & No tenant-visible NVLink fabric. & Scheduler assigns one confidential GPU. \\
Protected PCIe (Hopper) & Whole attached GPU complex & Fabric private to the tenant by assignment. & Scheduler sacrifices utilization to isolate a large unit. \\
Blackwell multi-GPU GPU-CC & 1/2/4/8-GPU NVSwitch partition & Shared fabric, host-managed, confidential partitions. & Scheduler allocates a fabric-valid partition shape. \\
\bottomrule
\end{longtable}
\normalsize
\endgroup

This paper measures the performance axis (Sections 4-6) on both a Blackwell
PCIe platform without NVLink and a B300 HGX, and the capability axis
(Section 7) on B300. NVLink/NVSwitch scale-up is in scope; InfiniBand/RDMA
scale-out, GPUDirect, and TEE-IO/TDISP device paths are future lanes
\citep{dmtf-spdm,pci-tdisp}.

\section{Platforms and Method}

\subsection{Platforms}

\begingroup
\scriptsize
\setlength{\tabcolsep}{2pt}
\setlength{\LTleft}{0pt}
\setlength{\LTright}{0pt}
\begin{longtable}{@{\extracolsep{\fill}}>{\RaggedRight\arraybackslash}p{0.054\linewidth}>{\RaggedRight\arraybackslash}p{0.158\linewidth}>{\RaggedRight\arraybackslash}p{0.158\linewidth}>{\RaggedRight\arraybackslash}p{0.257\linewidth}>{\RaggedRight\arraybackslash}p{0.274\linewidth}@{}}
\toprule
\textbf{GPU} & \textbf{Configuration} & \textbf{Fabric} & \textbf{Host / runtime} & \textbf{Role} \\
\midrule
\endfirsthead
\toprule
\textbf{GPU} & \textbf{Configuration} & \textbf{Fabric} & \textbf{Host / runtime} & \textbf{Role} \\
\midrule
\endhead
RTX Pro 6000 & 2x RTX Pro 6000 Blackwell Server Edition, 96 GB GDDR7 each & No NVLink; PCIe Gen5 & Ubuntu TDX stack (kernel 6.14-intel), dstack, CC-capable guest driver & Blackwell PCIe baseline; bridge microbenchmarks and loader/KV studies. \\
B300 HGX & 8x B300 SXM6 AC, 267 GiB HBM observed per GPU & NVLink/NVSwitch HGX fabric & Ubuntu 25.10, kernel 6.14.0-1008-intel, dstack 0.5.9, guest driver 595.58.03 & Flagship Blackwell server; serving, scheduling, and fabric experiments. \\
H200 & 1x H200 Hopper GPU & Single-GPU lane; NVLinks blocked in the CC-off comparison & Intel TDX/GPU-CC CVM paired with a plain-KVM CC-off VM on the same physical H200; vLLM 0.22.1 & Supplemental boundary check only; not a target for Blackwell performance or fabric-capability claims. \\
\bottomrule
\end{longtable}
\normalsize
\endgroup

The primary Blackwell hosts use the following Supermicro configurations:

\begingroup
\scriptsize
\setlength{\tabcolsep}{2pt}
\setlength{\LTleft}{0pt}
\setlength{\LTright}{0pt}
\begin{longtable}{@{\extracolsep{\fill}}>{\RaggedRight\arraybackslash}p{0.098\linewidth}>{\RaggedRight\arraybackslash}p{0.205\linewidth}>{\RaggedRight\arraybackslash}p{0.137\linewidth}>{\RaggedRight\arraybackslash}p{0.186\linewidth}>{\RaggedRight\arraybackslash}p{0.274\linewidth}@{}}
\toprule
\textbf{System} & \textbf{CPU} & \textbf{Memory} & \textbf{Local storage} & \textbf{Accelerators} \\
\midrule
\endfirsthead
\toprule
\textbf{System} & \textbf{CPU} & \textbf{Memory} & \textbf{Local storage} & \textbf{Accelerators} \\
\midrule
\endhead
Supermicro X14 2U CloudDC & 2x Intel Xeon 6730P, 32 cores each (64 cores/128 threads total) & 512 GB DDR5-4800 (8x 64 GB) & 2x Samsung 1.9 TB PCIe 5.0 NVMe; 2x Samsung 960 GB NVMe M.2 & 2x NVIDIA RTX PRO 6000 Blackwell Server Edition, 96 GB GDDR7 each \\
Supermicro X14 8U GPU System & 2x Intel Xeon 6787P, 86 cores each (172 cores/344 threads total) & 3,072 GB DDR5-5200 (32x 96 GB, 24-channel) & 1x Hynix 7.6 TB PCIe 5.0 NVMe; 2x Samsung 960 GB NVMe M.2 & NVIDIA HGX B300 8-GPU, 288 GB HBM3e per GPU (267 GiB observed by the benchmark stack) \\
\bottomrule
\end{longtable}
\normalsize
\endgroup

The two primary platforms answer different Blackwell questions. The Pro 6000
system isolates single-GPU PCIe behavior and supports long-running mechanism
studies. B300 adds the HGX dimensions: faster HBM, SXM packaging, and the
host-managed NVSwitch fabric that Section 7 qualifies. Cross-platform
agreement is itself a result: where the same software bottleneck produces the
same number on both Blackwell machines, the bottleneck is the confidential data
path, not the GPU.

H200 is included only as a boundary check. It uses the same microbenchmark and
Qwen3.6-27B-FP8 serving shapes to test whether the bridge law itself is unique
to Blackwell. These rows serve one purpose: paired CC-on/off runs on one
physical H200 reproduce the bridge microbenchmark law and show that CC removes
vLLM async scheduling's normal advantage on a Hopper GPU. The paper's
performance comparisons, capability claims, and operational conclusions remain
Blackwell results.

\subsection{Runtime Substrate}

All confidential VM experiments use dstack as the launch substrate: repeatable
CVM creation, GPU passthrough, and a common guest image family. Single-GPU
B300 CC runs used the standalone driver mode (\path|nvidia.NVreg_NvLinkDisable=1|
in the guest command line) to bypass a fabric-probe gate that otherwise fails
CUDA initialization on SXM parts; the repository README records the operational
details. Multi-GPU NVSwitch tenants use the fabric path instead; the two modes
are never mixed in one comparison.

\subsection{Experiment Families}

\begingroup
\scriptsize
\setlength{\tabcolsep}{2pt}
\setlength{\LTleft}{0pt}
\setlength{\LTright}{0pt}
\begin{longtable}{@{\extracolsep{\fill}}>{\RaggedRight\arraybackslash}p{0.211\linewidth}>{\RaggedRight\arraybackslash}p{0.323\linewidth}>{\RaggedRight\arraybackslash}p{0.366\linewidth}@{}}
\toprule
\textbf{Family} & \textbf{Controlled comparison} & \textbf{Question answered} \\
\midrule
\endfirsthead
\toprule
\textbf{Family} & \textbf{Controlled comparison} & \textbf{Question answered} \\
\midrule
\endhead
Bridge microbenchmarks & CC-off vs CC-on, same harness, same guest image & Which physical resource does CC tax, and by how much? \\
Serving + scheduling & CC-off vs CC-on, async vs sync vs patched scheduler, concurrency sweeps & Which runtime policies expose the bridge to users, and what recovers the loss? \\
Model loading & Loader variants on both platforms, CC-on & Does CC-aware movement engineering transfer across Blackwell platforms? \\
KV-cache offload & Offload policies, churn workloads, CC-off vs CC-on & How does CC change the economics of moving KV state? \\
Fabric qualification & Confidential tenants over NVSwitch partition shapes & What multi-GPU tenant shapes work, and what evidence can the tenant inspect? \\
\bottomrule
\end{longtable}
\normalsize
\endgroup

\subsection{Comparability and Claim Scope}

The B300 serving experiments used a newer \path|vllm/vllm-openai:latest| image
because the Pro 6000 vLLM build does not run on SM100. Cross-platform serving
rows are therefore capacity context and model-transfer evidence, not pure
hardware speedups; we flag this wherever it applies. Microbenchmarks and
loader runs do not share this caveat.

The B300 campaign was a time-bounded access window on hardware that has since
been returned. Headline numbers are preserved point measurements - single
benchmark executions, not statistically powered sweeps - and every one maps to
a raw artifact or preserved research note in the repository evidence index.
We use them to support large, qualitative effects (2x-44x mechanisms, 13-27\%
gaps, recovery of most of a gap) rather than small deltas. Where a result was
not reproduced before the window closed - notably the worker-thread scheduler
patch's B300 high-concurrency recovery number - we label the B300 number as
qualified and keep it out of the headline claims. The H200 port of the same
worker-thread idea confirms the mechanism, but not the exact B300 recovery
magnitude.

\section{The Serialized Bridge}

The microbenchmarks establish the mechanism that the rest of the paper
applies. The question is not "how much slower is CC?" but "which resource does
CC constrain, and what law governs it?"

\subsection{Compute and GPU-Local Memory Are at Parity}

BF16 matmul (8192x8192) on B300 runs at 1573.70 TFLOPS with CC off and
1569.81 TFLOPS with CC on - 0.998x. A single CUDA graph chaining 96,000
BF16 matmuls over 6.4 GB of resident weights, with zero CPU-GPU
synchronization between iterations, takes 7029.49 ms CC-off and 7037.80 ms
CC-on (1.0012x). A 1 GiB HBM-style access harness drops modestly to 0.912x.

\begingroup
\scriptsize
\setlength{\tabcolsep}{2pt}
\setlength{\LTleft}{0pt}
\setlength{\LTright}{0pt}
\begin{longtable}{@{\extracolsep{\fill}}>{\RaggedRight\arraybackslash}p{0.360\linewidth}>{\RaggedRight\arraybackslash}p{0.180\linewidth}>{\RaggedRight\arraybackslash}p{0.180\linewidth}>{\RaggedRight\arraybackslash}p{0.180\linewidth}@{}}
\toprule
\textbf{B300 microbenchmark} & \textbf{CC-off} & \textbf{CC-on} & \textbf{CC-on / CC-off} \\
\midrule
\endfirsthead
\toprule
\textbf{B300 microbenchmark} & \textbf{CC-off} & \textbf{CC-on} & \textbf{CC-on / CC-off} \\
\midrule
\endhead
BF16 matmul, 8192x8192 & 1573.70 TFLOPS & 1569.81 TFLOPS & 0.998x \\
96k chained matmuls, one CUDA graph & 7029.49 ms & 7037.80 ms & 1.0012x \\
HBM-style 1 GiB harness & 3156.06 GB/s & 2879.09 GB/s & 0.912x \\
Sustained H2D transfer, one context & 55.48 GB/s & 11.26 GB/s & 0.203x \\
Sustained D2H transfer, one context & 57.38 GB/s & 12.08 GB/s & 0.211x \\
Multiprocess H2D best & 55.55 GB/s & 34.18 GB/s & 0.615x \\
Multiprocess D2H best & 57.38 GB/s & 39.98 GB/s & 0.697x \\
\bottomrule
\end{longtable}
\normalsize
\endgroup

\begin{figure}[H]
\centering
\includegraphics[width=0.94\linewidth]{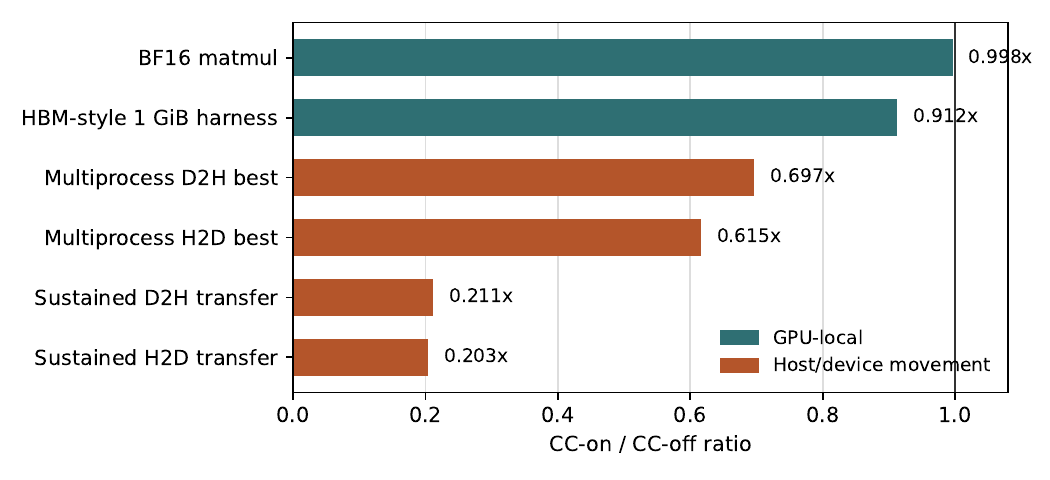}
\caption{B300 CC-on / CC-off ratio by microbenchmark.}
\end{figure}

Figure 1 shows the split: GPU-local work is at parity; bridge crossings fall
off a cliff. This is consistent with the architectural analysis of Gu et al.
\citep{gpu-cc-blueprint}: HBM is firewalled, not encrypted, so there is no per-byte
cipher on GPU-local traffic; only the PCIe staging path pays cryptographic and
staging costs.

\subsection{Context-Level, Not Stream-Level, Concurrency}

Sustained single-context host/device bandwidth drops to 0.20-0.21x under CC
on B300 (55.48 to 11.26 GB/s H2D), and to the same 11.5-11.7 GB/s level on
the Pro 6000 platform - the absolute floor is the secure copy path, not the
GPU generation. The H200 boundary experiment lands in the same range: 55.32
to 10.03 GB/s H2D and 55.14 to 10.35 GB/s D2H, with compute and HBM still
near parity.
This separates the law from its consequence. The secure-copy law is already
visible on Hopper; Blackwell amplifies it into larger serving penalties because
faster compute exposes the fixed bridge cost.

The recovery shape identifies the constrained resource. On Pro 6000 CC-on:

\begin{itemize}
\item \textbf{More streams in one context do nothing.} A multi-stream harness in a
single primary context stays at 4.9-5.0 GB/s H2D regardless of stream
count.
\item \textbf{More contexts scale.} Driver-API contexts within one process scale
aggregate H2D from about 5 GB/s at one context to about 35 GB/s at 24;
separate processes scale the same way (10 to 35 GB/s). B300 reproduces the
multiprocess recovery (0.62-0.70x at best).
\end{itemize}

\begin{figure}[H]
\centering
\includegraphics[width=0.94\linewidth]{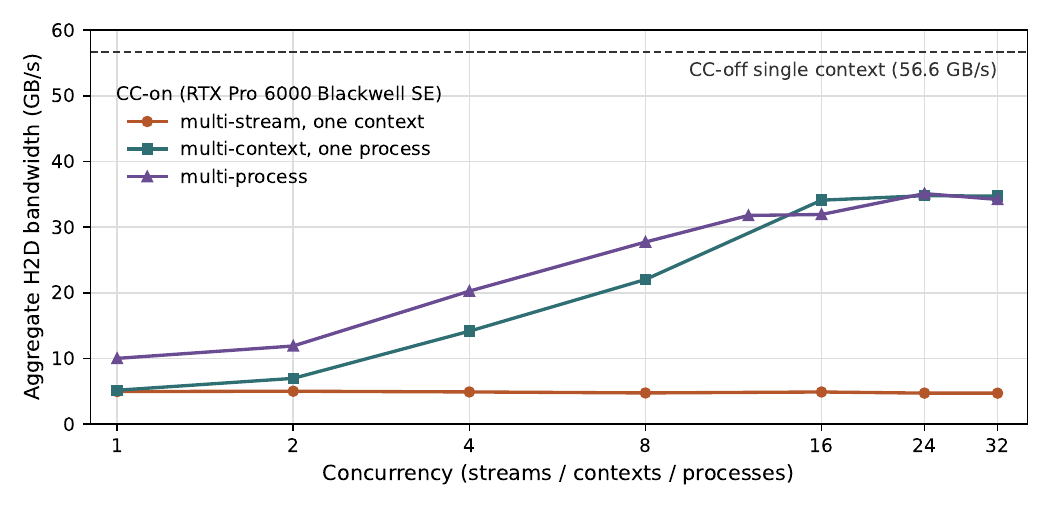}
\caption{CC-on transfer bandwidth vs concurrency: streams flat, contexts scale.}
\end{figure}

Figure 2 makes the law visible: the secure copy channel is a per-context
resource, and within a context all traffic serializes. NVIDIA's operations
guide says as much for context limits \citep{nvidia-secops}; the measurement shows
the performance consequence.

Two finer-grained probes complete the picture:

\begin{itemize}
\item \textbf{Small-copy serialization.} Sixteen CUDA streams each issuing 32-byte D2H
copies achieve approximately zero parallelism CC-on (40 microseconds per
copy single-stream, 39 at sixteen streams) versus about 24\% scaling CC-off
(17 to 13 microseconds). Under CC the hardware genuinely cannot overlap
same-context copies.
\item \textbf{Asynchrony is revoked.} \path|dst.copy_(src, non_blocking=True)| with a
pinned destination still blocks the calling CPU thread for the full
transfer under CC. The driver cannot return before bounce-buffer setup
completes. We measure the setup cost at roughly 330 microseconds per call
(Section 5.2) - for a 32-byte payload, three orders of magnitude above the
per-byte cost.
\end{itemize}

The same H200 experiment repeats the stream and small-copy signatures with
different absolute timings but the same shape: 32-byte D2H copies stay flat
from one to sixteen streams under CC (35 to 34 microseconds), and multiple
contexts recover bandwidth where streams do not. That rules out a B300
software-stack artifact and leaves the Blackwell-specific question at the
serving layer: why the same fixed bridge cost grows into larger penalties on
faster hardware.

\subsection{Cipher Throughput Is Not the Limiter}

Two ablations rule out raw cryptographic throughput as the limiter. Disabling
AES-NI and PCLMUL in the guest's OpenSSL collapses CC-on duplex bandwidth from
40.4 to 5.5 GB/s - so CPU AES-GCM is causally on the path. But disabling only
the wider-vector VAES/VPCLMUL implementations costs just 3.4\% - so the path is
not compute-bound on cipher width at the observed plateau. Combined with the
context-scaling result, the bandwidth ceiling is set by the secure-channel
architecture and its per-crossing costs, not by how fast the CPU can encrypt.

\subsection{Bridge Model Summary}

The bridge behaves as a serialized, high-setup-cost channel:

\begin{enumerate}
\item Within a CUDA context, cross-device transfers serialize on a fixed pool of
secure copy channels; stream-level overlap is a fiction under CC.
\item Each crossing pays a fixed setup toll (about 330 microseconds observed),
so many small crossings are catastrophically worse than few large ones.
\item Additional bandwidth requires additional CUDA contexts, each with its own
expensive secure lifecycle (Section 6.1 measures 5.2 s of context creation
for eight workers).
\item GPU-local compute and memory are at parity; only crossings are taxed.
\end{enumerate}

Everything that follows - the Blackwell scheduling inversion, the loader
design, the KV policy - is this law applied to a different layer of the
serving stack.

\section{Case Study: Policy Inversion in the Serving Runtime}

\subsection{Workload-Dependent Serving Overheads}

Across preserved B300 serving rows, the CC tax ranges from negligible to
catastrophic depending on how often the workload crosses the bridge:

\begingroup
\scriptsize
\setlength{\tabcolsep}{2pt}
\setlength{\LTleft}{0pt}
\setlength{\LTright}{0pt}
\begin{longtable}{@{\extracolsep{\fill}}>{\RaggedRight\arraybackslash}p{0.455\linewidth}>{\RaggedRight\arraybackslash}p{0.167\linewidth}>{\RaggedRight\arraybackslash}p{0.139\linewidth}>{\RaggedRight\arraybackslash}p{0.139\linewidth}@{}}
\toprule
\textbf{Workload class} & \textbf{B300 CC-off} & \textbf{B300 CC-on} & \textbf{Delta} \\
\midrule
\endfirsthead
\toprule
\textbf{Workload class} & \textbf{B300 CC-off} & \textbf{B300 CC-on} & \textbf{Delta} \\
\midrule
\endhead
MLPerf-shaped online serving, GPT-OSS-120B, qps=1.19 & 1193 tok/s & 1180 tok/s & -1.1\% \\
Dense decode, Qwen3.6-27B-FP8 (serving matrix row) & 3302 tok/s & 2873 tok/s & -13.0\% \\
Dense decode, Gemma-4-31B-it, c=128 & 2357 tok/s & 2022 tok/s & -14.2\% \\
MoE decode, Qwen3.6-35B-A3B-FP8 & 5282 tok/s & 3981 tok/s & -24.7\% \\
MoE decode, Gemma-4-26B-A4B-it & 5583 tok/s & 4040 tok/s & -27.6\% \\
KV restore, Qwen2.5-3B churn=3 warm TTFT & 405.4 ms & 935.2 ms & +131\% \\
\bottomrule
\end{longtable}
\normalsize
\endgroup

\begin{figure}[H]
\centering
\includegraphics[width=0.94\linewidth]{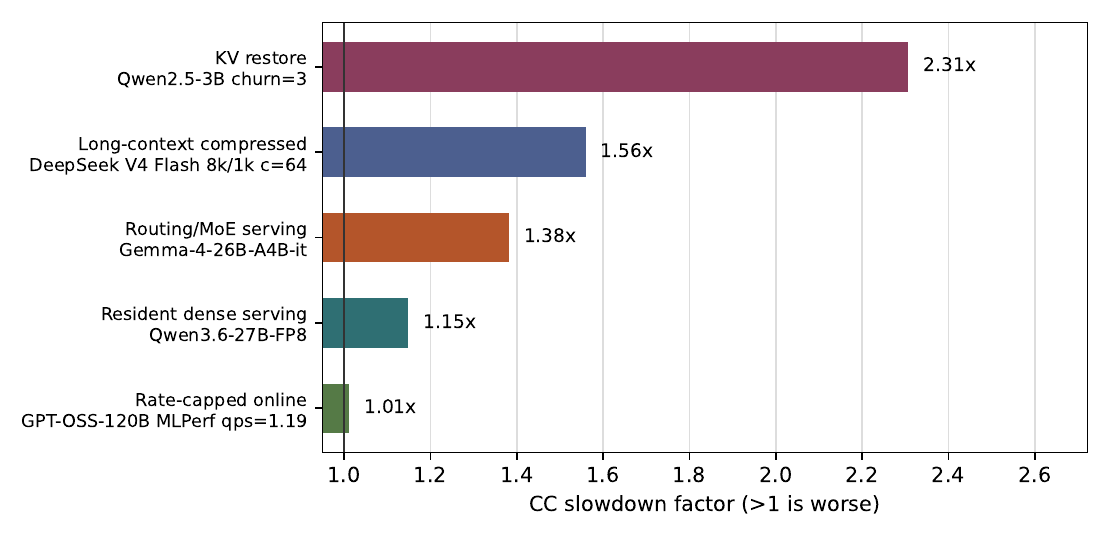}
\caption{B300 serving slowdown by workload class.}
\end{figure}

Figure 3 normalizes representative rows to a slowdown factor. Rate-capped
online serving hides the tax (the bridge is idle most of the time); resident
dense decode pays a moderate tax; MoE pays more (more frequent small
cross-device operations per step); KV restore - bulk CPU-to-GPU movement on
the critical path - is the extreme. Deployment-style one-GPU rows show the
same spread: DeepSeek V4 Flash at 8k/1k loses 1.56-1.69x and
GLM-4.7-Flash-MTP-NVFP4 loses 1.30-2.42x depending on concurrency, with the
single-stream, speculative-decoding-heavy configuration worst. A single "CC
overhead percentage" does not exist; the tax is a function of bridge-crossing
frequency and size.

Two cross-platform observations sharpen the question. First, B300 delivers
1.55-3.80x the absolute CC-on throughput of Pro 6000 on matched rows (newer
vLLM image required on SM100; Section 3.4), so confidential serving capacity
clearly scales with hardware. Second, the \emph{relative} tax grows with faster
hardware: the same stack and workload that loses about 10\% on Pro 6000 loses
about 26\% on B300 in the profiling configuration below, and KV-restore
degradation worsens from +43\% to +131\% even as absolute latency improves
1.74x. Faster GPUs shrink everything except the bridge.

\subsection{Accounting for the Dense-Decode Gap}

We profile the worst well-controlled case: Qwen3.6-27B-FP8, single B300 GPU,
128 prompts at concurrency 128 with 1k/1k token shapes, vLLM \texttt{:latest} with
CUDA graphs on, warm steady state at 5070 tok/s CC-off vs 3729 CC-on (-26\%,
TPOT 21.5 vs 30.2 ms).

The GPU is exonerated first: the 96k-chained-matmul graph (Section 4.1) shows
1.0012x on this exact machine pair, and eager-mode probes slow CC-off and
CC-on equally, so neither kernels nor launch overheads explain the gap. (The
profiled configuration differs from the Section 5.1 serving-matrix row and
isolates the warm steady-state window, where the relative gap is larger than
a whole-bench average that includes startup effects; this is why the same
model shows -13\% in the matrix and -26\% here.)

PyTorch profiling under both modes attributes the slowdown to one operation
class. Grouping \path|aten::copy_| calls by parent:

\begingroup
\scriptsize
\setlength{\tabcolsep}{2pt}
\setlength{\LTleft}{0pt}
\setlength{\LTright}{0pt}
\begin{longtable}{@{\extracolsep{\fill}}>{\RaggedRight\arraybackslash}p{0.346\linewidth}>{\RaggedRight\arraybackslash}p{0.115\linewidth}>{\RaggedRight\arraybackslash}p{0.138\linewidth}>{\RaggedRight\arraybackslash}p{0.115\linewidth}>{\RaggedRight\arraybackslash}p{0.185\linewidth}@{}}
\toprule
\textbf{Parent op} & \textbf{Calls} & \textbf{CC-off avg} & \textbf{CC-on avg} & \textbf{Per-call slowdown} \\
\midrule
\endfirsthead
\toprule
\textbf{Parent op} & \textbf{Calls} & \textbf{CC-off avg} & \textbf{CC-on avg} & \textbf{Per-call slowdown} \\
\midrule
\endhead
\path|aten::_to_copy| (alloc + H2D) & 1138 & 31.7 us & 1389 us & 44x \\
\path|copy_| into pre-allocated tensor & 2628 & 25.1 us & 31.0 us & 1.2x \\
\path|_prepare_inputs| pinned helper & 260 & 18.2 us & 18.4 us & 1.0x \\
Attention-path copies & 192 & 27.0 us & 27.8 us & 1.0x \\
\bottomrule
\end{longtable}
\normalsize
\endgroup

\path|aten::_to_copy| - PyTorch's fresh pinned allocation plus device-transfer path
\begin{itemize}
\item accounts for +1545 ms of the +1561 ms total slowdown. The accounting
\end{itemize}
closes: 1,357 microseconds of per-call delta times 1,138 calls is 1.54 s,
which is the observed 26\% throughput drop. The call sites are vLLM's per-step
scatter-index and sampling-index tensors: six small fresh-pinned H2D copies
per decode step.

This is the bridge law at the application layer: many small crossings, each
paying the roughly 330 microsecond bounce-buffer setup toll, allocated fresh
each step so no staging is reused, issued \path|non_blocking| under the assumption
they overlap - which CC revokes.

\subsection{Patch-Based Refutation of Alternative Causes}

Before changing policy, we tested and rejected the local explanations with
single-variable vLLM patches, all on the same workload:

\begingroup
\scriptsize
\setlength{\tabcolsep}{2pt}
\setlength{\LTleft}{0pt}
\setlength{\LTright}{0pt}
\begin{longtable}{@{\extracolsep{\fill}}>{\RaggedRight\arraybackslash}p{0.275\linewidth}>{\RaggedRight\arraybackslash}p{0.400\linewidth}>{\RaggedRight\arraybackslash}p{0.225\linewidth}@{}}
\toprule
\textbf{Hypothesis} & \textbf{Patch} & \textbf{Result} \\
\midrule
\endfirsthead
\toprule
\textbf{Hypothesis} & \textbf{Patch} & \textbf{Result} \\
\midrule
\endhead
Small H2D batching alone fixes it & v1: scatter-index batch 2-to-1 & -1.0 ms TPOT; capped \\
Fresh pinned allocation is the cost & v4: persistent pinned destination & +0.7 ms; no win \\
Extra stream / \path|wait_stream| overhead & v5: remove both & -0.7 ms; no win \\
Async-path setup ops in \path|__init__| & v8: persistent buffers, event ring, default stream & -0.5 ms; no win \\
CUDA-graph mode interacts & v9: FULL capture & slightly worse than PIECEWISE \\
\bottomrule
\end{longtable}
\normalsize
\endgroup

The structural patches all left TPOT at 30-31 ms. The only change that moved
it was scheduling policy. vLLM's default \texttt{--async-scheduling} overlaps step N's
device-to-host output drain with step N+1's preparation and launches:

\begingroup
\scriptsize
\setlength{\tabcolsep}{2pt}
\setlength{\LTleft}{0pt}
\setlength{\LTright}{0pt}
\begin{longtable}{@{\extracolsep{\fill}}>{\RaggedRight\arraybackslash}p{0.248\linewidth}>{\RaggedRight\arraybackslash}p{0.248\linewidth}>{\RaggedRight\arraybackslash}p{0.405\linewidth}@{}}
\toprule
\textbf{} & \textbf{CC-off} & \textbf{CC-on} \\
\midrule
\endfirsthead
\toprule
\textbf{} & \textbf{CC-off} & \textbf{CC-on} \\
\midrule
\endhead
DMA concurrency available & abundant & one secure channel at a time \\
The intended overlap & real parallelism & fiction - transfers serialize anyway \\
Overhead of attempting it & about zero & stream setup + arbitration between in-flight transfers \\
Net per step & about +3 ms saved & about -4 ms lost \\
\bottomrule
\end{longtable}
\normalsize
\endgroup

This is policy inversion: the optimization's entire benefit assumes concurrent
DMA, the one property the secure bridge removes, while its overhead remains.
H200 provides the boundary case. Without CC, async also helps there (3497
tok/s versus 3174 tok/s sync, +10\%). With CC, async and sync tie (3106 versus
3133 tok/s). This contrast matters: Hopper shows neutralization, while
Blackwell shows inversion, where a default optimization becomes the wrong
policy.

\subsection{Recovery with Synchronous Scheduling}

Disabling async scheduling converts the pattern to forward, sample, one small
D2H (about 512 bytes of sampled tokens), drain, continue - exactly the
sequential, drained pattern the bridge is engineered for:

\begingroup
\scriptsize
\setlength{\tabcolsep}{2pt}
\setlength{\LTleft}{0pt}
\setlength{\LTright}{0pt}
\begin{longtable}{@{\extracolsep{\fill}}>{\RaggedRight\arraybackslash}p{0.450\linewidth}>{\RaggedRight\arraybackslash}p{0.300\linewidth}>{\RaggedRight\arraybackslash}p{0.150\linewidth}@{}}
\toprule
\textbf{Qwen3.6-27B-FP8, c=128} & \textbf{Throughput} & \textbf{TPOT} \\
\midrule
\endfirsthead
\toprule
\textbf{Qwen3.6-27B-FP8, c=128} & \textbf{Throughput} & \textbf{TPOT} \\
\midrule
\endhead
CC-off async (gold) & 4522 tok/s & 23.64 ms \\
CC-on async (default) & 3550 tok/s & 31.10 ms \\
CC-off sync & 4153 tok/s & 26.56 ms \\
CC-on sync & 4104 tok/s & 26.92 ms \\
\bottomrule
\end{longtable}
\normalsize
\endgroup

The flag closes 57\% of the CC gap. More telling is the residual: CC-on sync
vs CC-off sync differ by 0.36 ms TPOT - about 1\%, within noise, confirmed by
per-phase instrumentation showing single-digit-microsecond to
sub-millisecond deltas in every engine phase. Under a bridge-respecting
schedule, the CC tax on dense decode essentially vanishes. The remaining gap
to gold (26.92 vs 23.64 ms) is the ordinary cost of synchronous scheduling -
CC-off pays the same 2.92 ms for it - not a confidential-computing cost.

The same flag generalizes by traffic pattern, as the law predicts:

\begingroup
\scriptsize
\setlength{\tabcolsep}{2pt}
\setlength{\LTleft}{0pt}
\setlength{\LTright}{0pt}
\begin{longtable}{@{\extracolsep{\fill}}>{\RaggedRight\arraybackslash}p{0.314\linewidth}>{\RaggedRight\arraybackslash}p{0.105\linewidth}>{\RaggedRight\arraybackslash}p{0.105\linewidth}>{\RaggedRight\arraybackslash}p{0.209\linewidth}>{\RaggedRight\arraybackslash}p{0.167\linewidth}@{}}
\toprule
\textbf{Workload} & \textbf{CC-on async} & \textbf{CC-on sync} & \textbf{Async-mode CC penalty recovered} & \textbf{Residual CC tax under sync} \\
\midrule
\endfirsthead
\toprule
\textbf{Workload} & \textbf{CC-on async} & \textbf{CC-on sync} & \textbf{Async-mode CC penalty recovered} & \textbf{Residual CC tax under sync} \\
\midrule
\endhead
Dense decode (Qwen3.6-27B) & 3550 tok/s & 4104 tok/s & \textasciitilde{}100\% & \textasciitilde{}1\% \\
KV-churn warm TTFT (Qwen2.5-3B) & 935 ms & 413 ms & \textasciitilde{}100\% & \textasciitilde{}2\% \\
MoE decode (Qwen3.6-35B-A3B) & 3981 tok/s & 4296 tok/s & \textasciitilde{}80\% & \textasciitilde{}5\% \\
\bottomrule
\end{longtable}
\normalsize
\endgroup

KV churn recovers completely because its async-mode loss was bulk transfers
contending with scheduling traffic on one channel. MoE retains about 5\%
because expert routing adds small, frequent, per-step crossings that no
schedule can remove - irreducible bridge traffic at the framework level.

\subsection{Recovery with a Worker-Thread Drain}

Synchronous scheduling forfeits real overlap that CC-off enjoys. Concurrency
sweeps on fresh CVM deployments show what is left on the table: CC-on async
ties CC-on sync at every concurrency (5026 vs 5004 tok/s at c=512 - the
pipelining benefit is fully destroyed), while CC-off async beats CC-off sync
by 15\%.

The bridge law says the CPU thread blocks because the driver cannot return
before staging setup; but a blocked CUDA call releases the GIL. Moving the
blocking drain to a Python worker thread therefore restores pipelining
without asking the hardware to overlap copies: the engine's main thread
submits step N+1's preparation and forward launches while the worker sits in
the driver. The patch (v10c) is about 30 lines in vLLM's async output path.

\begingroup
\scriptsize
\setlength{\tabcolsep}{2pt}
\setlength{\LTleft}{0pt}
\setlength{\LTright}{0pt}
\begin{longtable}{@{\extracolsep{\fill}}>{\RaggedRight\arraybackslash}p{0.100\linewidth}>{\RaggedRight\arraybackslash}p{0.175\linewidth}>{\RaggedRight\arraybackslash}p{0.125\linewidth}>{\RaggedRight\arraybackslash}p{0.125\linewidth}>{\RaggedRight\arraybackslash}p{0.150\linewidth}>{\RaggedRight\arraybackslash}p{0.225\linewidth}@{}}
\toprule
\textbf{c} & \textbf{CC-on vanilla async} & \textbf{CC-on sync} & \textbf{CC-on v10c} & \textbf{CC-off async (gold)} & \textbf{v10c gap to gold} \\
\midrule
\endfirsthead
\toprule
\textbf{c} & \textbf{CC-on vanilla async} & \textbf{CC-on sync} & \textbf{CC-on v10c} & \textbf{CC-off async (gold)} & \textbf{v10c gap to gold} \\
\midrule
\endhead
128 & 3629 & 3856 & 3942 & 4653 & -15.3\% \\
256 & - & 4766 & 5073 & - & -13.4\% (vs estimated gold) \\
512 & 5026 & 5004 & 5518 & 6020 & -8.3\% \\
\bottomrule
\end{longtable}
\normalsize
\endgroup

\begin{figure}[H]
\centering
\includegraphics[width=0.94\linewidth]{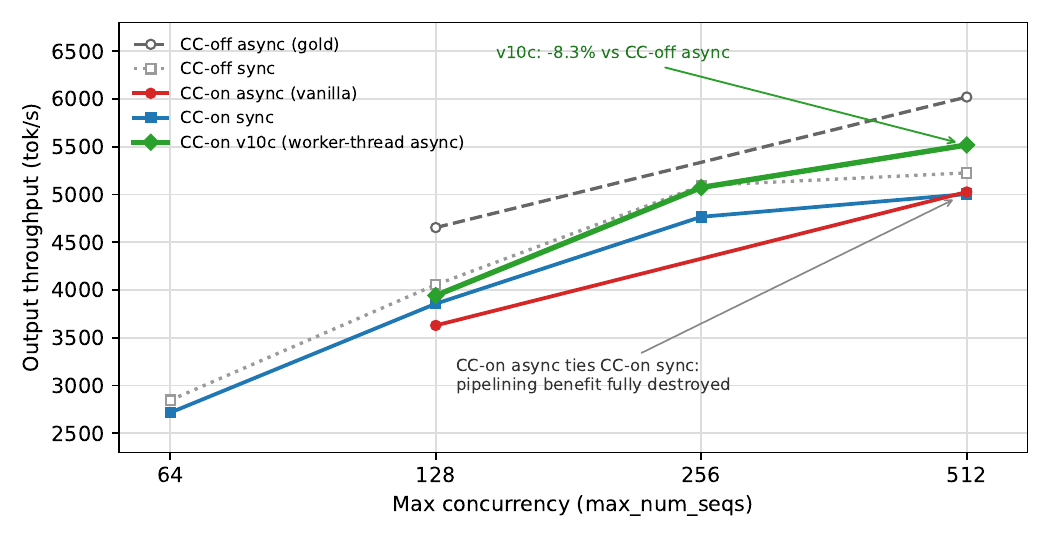}
\caption{Scheduling modes vs concurrency under GPU-CC on B300.}
\end{figure}

Figure 4 is the inversion and its repair in one plot: the default async path
is the worst CC-on configuration at c=128; the one-flag sync fix recovers
most of the gap; and v10c recovers half of what remains, reaching within
8.3\% of the non-confidential gold configuration at c=512 - while \emph{beating}
CC-off sync (5226 tok/s) at the same point. The residual is the GPU-side
stream pipelining that the secure-channel architecture removes regardless of
host structure; closing it requires multi-step CUDA graphs or a trusted
device path (TDISP).

Because the B300 window closed before an independent rerun, we label v10c a
qualified result: the patch source, run logs, and sweep notes are preserved
(in the repository evidence index), but the headline recovery claims of this
paper rest on the one-flag fix, which was reproduced across workloads. The
same worker-thread idea was also ported to vLLM 0.22.1 on H200, improving
CC-on throughput by 6-7\% across c=64-512. That confirms the mechanism - moving
a blocking drain off the engine thread helps - while leaving the B300
high-concurrency recovery number qualified.

The recovery hierarchy, in deployment terms:

\begingroup
\scriptsize
\setlength{\tabcolsep}{2pt}
\setlength{\LTleft}{0pt}
\setlength{\LTright}{0pt}
\begin{longtable}{@{\extracolsep{\fill}}>{\RaggedRight\arraybackslash}p{0.234\linewidth}>{\RaggedRight\arraybackslash}p{0.288\linewidth}>{\RaggedRight\arraybackslash}p{0.378\linewidth}@{}}
\toprule
\textbf{Effort} & \textbf{CC gap recovered} & \textbf{Step} \\
\midrule
\endfirsthead
\toprule
\textbf{Effort} & \textbf{CC gap recovered} & \textbf{Step} \\
\midrule
\endhead
One CLI flag & 57\% (dense), \textasciitilde{}100\% (KV churn) & \texttt{--no-async-scheduling} \\
\textasciitilde{}30-line runtime patch & B300: up to 92\% at c=512 (qualified); H200: +6-7\% & worker-thread drain (v10c) \\
Architectural & most of the rest & multi-step CUDA graphs \\
Platform & bridge cost itself & TDISP / TEE-IO trusted device path \\
\bottomrule
\end{longtable}
\normalsize
\endgroup

\subsection{Runtime Design Rule}

GPU-CC's transfer machinery is engineered for sequential, drained transfers;
the modern accelerator software idiom - saturate the device with overlapped
streams and asynchronous copies - is its adversarial input. Any framework
that pipelines D2H/H2D across streams expecting parallel DMA will pay an
outsized CC penalty, and the fix shape is general: drain before refill, or
move blocking waits off the critical thread. Candidates include multimodal
input pipelines, parameter-sharded inference, asynchronous tokenization, and
RAG pipelines. A CC-aware runtime should treat bridge crossings as a
scheduled, scarce resource - batched, drained, and kept off the critical
path - the runtime analogue of the fabric result in Section 7, where another
implicit resource becomes an explicit scheduling object.

\section{Case Study: Movement Engineering for Loading and KV State}

\subsection{Context-Pooled Model Loading}

Loading GPT-OSS-120B (59 GiB, 15 shards) into a confidential GPU with vLLM's
default safetensors path takes 287.09 s on B300 and 287.41 s on Pro 6000.
The near-identical numbers on hardware a generation of memory bandwidth apart
are the tell: the bottleneck is the software path - serialized parsing,
staging, and single-context secure transfer - not the GPU.

The fix applies the bridge law directly. Bandwidth lives in contexts
(Section 4.2), so the loader fans weight shards across a pool of worker CUDA
contexts that stage host-to-device copies in parallel, then assembles into
the framework's tensors with same-GPU peer copies (26.8-32.7 GB/s observed
on Pro 6000). But secure contexts are expensive: for eight workers we measure
5.20 s of \texttt{cuCtxCreate}, 3.90 s of \texttt{cuCtxDestroy}, and 0.30 s of pinned-slot
allocation - 9.4 s of lifecycle cost per load if paid on the critical path,
which is why a naive per-shard-context variant takes 253.66 s. Persistent
context pooling amortizes it (19.99 s); prewarming the pool before the weight
iterator starts and tearing it down asynchronously after removes it from the
critical path entirely.

\begingroup
\scriptsize
\setlength{\tabcolsep}{2pt}
\setlength{\LTleft}{0pt}
\setlength{\LTright}{0pt}
\begin{longtable}{@{\extracolsep{\fill}}>{\RaggedRight\arraybackslash}p{0.500\linewidth}>{\RaggedRight\arraybackslash}p{0.200\linewidth}>{\RaggedRight\arraybackslash}p{0.200\linewidth}@{}}
\toprule
\textbf{GPT-OSS-120B loader path} & \textbf{B300 CC-on} & \textbf{Pro 6000 CC-on} \\
\midrule
\endfirsthead
\toprule
\textbf{GPT-OSS-120B loader path} & \textbf{B300 CC-on} & \textbf{Pro 6000 CC-on} \\
\midrule
\endhead
safetensors baseline & 287.09 s & 287.41 s \\
safetensors, 8 threads & 56.82 s & 66.79 s \\
fastsafetensors \citep{fastsafetensors} & 36.34 s & 36.83 s \\
+ CC-aware context pool, persistent 8 workers & 19.99 s & 20.46 s \\
+ prewarm and async teardown (best, warm) & 8.36 s & 8.80 s \\
\bottomrule
\end{longtable}
\normalsize
\endgroup

\begin{figure}[H]
\centering
\includegraphics[width=0.94\linewidth]{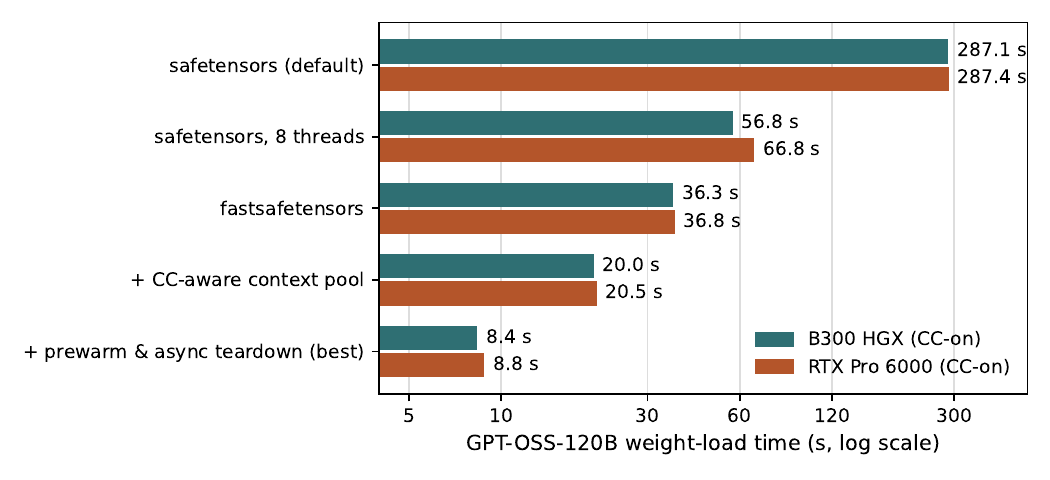}
\caption{GPT-OSS-120B weight-load time by loader variant on both Blackwell platforms.}
\end{figure}

Figure 5 shows every optimization stage transferring across platforms within
5\%. This matters operationally: model load time is tenant cold-start, model
switching, and failure recovery. At 287 s a scheduler cannot treat
confidential GPU capacity as elastic; at 8.4 s it can. It also matters
architecturally: the loader is the existence proof that the bridge law is
actionable - the same data, moved with CC-appropriate concurrency (contexts,
pooled lifecycles, large drained transfers), approaches the multiprocess
bandwidth ceiling of Section 4.2.

\subsection{Reuse-Aware KV-Cache Offload}

KV restore is the movement-heavy extreme of serving (Figure 3): restoring a
1.1 GiB prefix per request under churn raises warm TTFT by +131\% on B300
(405 to 935 ms) and +43\% on Pro 6000. Two levers recover it:

\begin{itemize}
\item \textbf{Schedule (Section 5.4):} sync scheduling alone brings B300 churn TTFT
from 935 to 413 ms - the bulk restore stops contending with per-step
scheduling traffic for the serialized channel.
\item \textbf{Policy:} the default vLLM CPU-offload path spills far more than it
restores (multi-GiB device-to-host against MiB-scale host-to-device in the
measured configuration). Reuse-aware filtering (\path|store_threshold=2|, only
offload blocks observed at least twice) cuts spill volume from 2.3 GiB to
2.3 MB and improves CC-on warm TTFT 2.97x (1615 to 544 ms) on Pro 6000.
\end{itemize}

The principle generalizes the loader result: under CC, every byte moved
across the bridge costs more, so policies that move bytes speculatively -
spill-everything offload, eager prefix restore, fine-grained streaming -
must become evidence-driven. Residency is worth more under CC than off it,
and Blackwell's larger HBM makes residency cheaper to buy.

\section{Blackwell Capability: Confidential Multi-GPU Fabric Tenants}

B300 HGX changes what the scheduled object is. On PCIe platforms, a
confidential tenant owns passed-through devices. On B300, a tenant receives a
\emph{partition of a shared NVSwitch fabric} - on this platform, eight GPUs behind
two QM3 NVSwitches reached over CX-7 management NICs, versus four NVSwitches
on H100/H200 HGX - programmed by host-side Fabric Manager (FM) and NVLSM
before tenant boot \citep{nvidia-fabric-manager,nvidia-shared-nvswitch}. This
section reports what we qualified during the access window - to our knowledge
the first public measurements of this mode - and what trust questions it
opens.

\begin{figure}[H]
\centering
\includegraphics[width=0.94\linewidth]{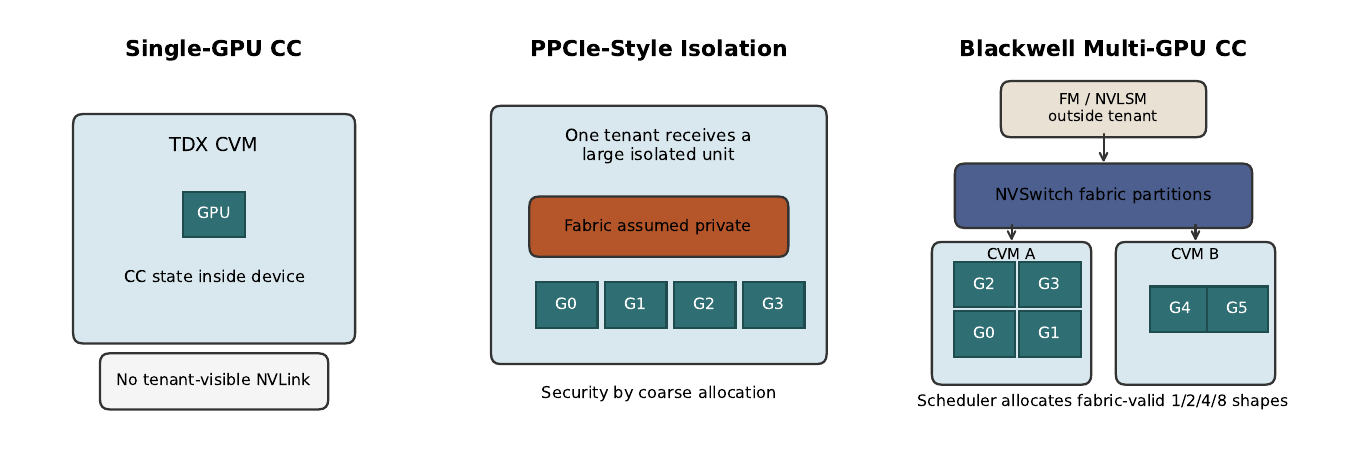}
\caption{Confidential GPU modes and fabric ownership.}
\end{figure}

\subsection{Validated Fabric Capabilities}

\begingroup
\scriptsize
\setlength{\tabcolsep}{2pt}
\setlength{\LTleft}{0pt}
\setlength{\LTright}{0pt}
\begin{longtable}{@{\extracolsep{\fill}}>{\RaggedRight\arraybackslash}p{0.105\linewidth}>{\RaggedRight\arraybackslash}p{0.795\linewidth}@{}}
\toprule
\textbf{Capability} & \textbf{Evidence} \\
\midrule
\endfirsthead
\toprule
\textbf{Capability} & \textbf{Evidence} \\
\midrule
\endhead
Partition vocabulary & FM enumerates 15 partition definitions on this system: one 8-GPU, two 4-GPU, four 2-GPU, eight 1-GPU shapes. Activation/deactivation (\texttt{fmpm -a}/\texttt{-d}) takes 10-20 s per tenant. \\
n=2 confidential NVLink tenant & CUDA initializes, peer access works both directions, fabric reports completed/healthy, the guest driver reports NVLink-encryption (NVLE) multi-GPU mode, and \texttt{cudaMemcpyPeer} sustains 510.4 GB/s over NVLink inside the CVM. \\
n=4 confidential tenant & All four GPUs report healthy fabric state with a full NV18 mesh topology inside the CVM; CUDA workload validation was deferred by a container-toolkit issue, not an architectural one. \\
Concurrent tenant isolation & Two simultaneous n=2 confidential tenants on disjoint partitions each see exactly their two GPUs, both report CC on and healthy NV18 fabric, with no host management NICs exposed to either tenant. In an earlier qualification lane, two simultaneous n=4 confidential tenants also booted on disjoint partitions with NVLink-encryption mode, full NV18 mesh, and passing GPU attestation on both; CUDA workloads were gated at that date by a since-resolved guest-userspace issue. \\
\bottomrule
\end{longtable}
\normalsize
\endgroup

The n=2 bandwidth point deserves emphasis: 510 GB/s of GPU-to-GPU bandwidth
\emph{inside a confidential tenant} is two orders of magnitude above what the
CVM-GPU bridge sustains (Section 4.2), and it does not transit host memory.
The scale-up fabric is thus the one data path GPU-CC does not serialize -
which makes fabric tenancy a performance feature, not only a capacity
feature. (The contrast is stark in failure: with NVLink disabled, NCCL falls
back to a CC-compatible TCP path measured at about 10 MB/s.)

What we did not qualify: n=8 CUDA/P2P/NCCL (blocked by a host FM
cache-staleness bug during the window; topology evidence only), NCCL
collectives at n=2/4 (deferred), and any CC-on vs CC-off NVLink overhead
comparison. These remain open measurements, not open capabilities - NVIDIA
documents 1/2/4/8 shapes as supported.

\subsection{Scheduling Consequences}

The partition vocabulary is the scheduling API. A B300-class control plane
allocates fabric-valid shapes (1/2/4/8), not arbitrary GPU sets; partitions
are activated per-tenant against a shared fabric, so utilization no longer
requires Protected-PCIe-style whole-complex assignment. dstack drives this
today: partition activation, CVM launch with the partition's GPUs, health
gates on FM state before boot. The operational failure modes we hit - stale
FM partition state surfacing as guest FLA remap validation errors - argue for
fabric-state health checks as a scheduling precondition, exactly as NUMA
topology became a first-class scheduler input a decade ago, but with a trust
dimension NUMA never had.

\subsection{Remaining Fabric-Attestation Gap}

The tenant can verify: TDX evidence for its CVM, GPU-CC mode and ready state,
GPU attestation reports, and guest-visible fabric health and NVLink topology.
The tenant cannot verify: the Fabric Manager binary and configuration that
programmed its partition, or the NVSwitch routing tables that decide where
its NVLink traffic can go. Today that control plane is host-trusted. The
fabric works as a confidential resource before it is an \emph{attestable} one.

Closing the gap has a plausible path - moving FM into an attestable service
VM, signing routing state, and eventually TDISP-style device interface
attestation \citep{pci-tdisp} - but until then, multi-GPU confidential tenants on
B300 carry a documented host-control-plane dependency that single-GPU
PCIe tenants do not. Platforms should state it rather than paper over it.

\section{Design Principles for Confidential AI Platforms}

The measurements compress into five rules:

\begin{enumerate}
\item \textbf{Treat bridge crossings as a scheduled resource.} Batch small transfers,
drain before refill, and keep crossings off the critical path. The
difference between honoring and ignoring this rule is the difference
between a 1\% and a 26\% serving tax (Section 5).
\item \textbf{Get concurrency from contexts, pooled.} Stream-level overlap is void
under CC; context-level parallelism works but its lifecycle is expensive.
Pool and prewarm (Section 6.1).
\item \textbf{Make policy defaults CC-mode-aware.} The best default without CC can be
the worst with it. Runtimes should detect GPU-CC mode and flip scheduling,
offload, and streaming defaults accordingly - today that is a deployment
checklist (\texttt{--no-async-scheduling}; reuse-aware offload thresholds);
properly it belongs in the runtime.
\item \textbf{Buy residency.} Weights and KV state that stay on the GPU are free;
everything that moves is taxed. Larger Blackwell HBM raises the return on
residency-preserving policies.
\item \textbf{Schedule and attest the fabric.} On B300-class systems the tenant shape
is a fabric partition. Platforms should expose partition identity, fabric
health, and GPU-CC state to tenants now, and drive toward attestable
fabric control (Section 7.3).
\end{enumerate}

TEE-IO/TDISP sits at the end of this road: a trusted device path would shrink
the bridge toll itself, converting rules 1-2 from necessities into
optimizations. Until it ships in mainline stacks, the recoveries in this
paper are what production confidential serving has.

\section{Related Work}

\textbf{GPU TEEs before commercial GPU-CC.} Graviton proposed GPU trusted
execution with secure command submission and protected copies \citep{graviton};
Telekine moved GPU TEE interaction behind an API-remoting boundary
\citep{telekine}. These works anticipated that the host-device path, not device
compute, is where accelerator security costs concentrate - the property we
measure in production silicon.

\textbf{NVIDIA GPU-CC characterization.} NVIDIA's H100 material describes the
confidential mode and its encrypted staging design [NVIDIA-H100-CC;
NVIDIA-SECURE-AI; NVIDIA-SECOPS]. Gu et al. reconstruct the GPU-CC
architecture, boot, and runtime bridge in detail \citep{gpu-cc-blueprint}; our
contribution is quantifying the performance law that bridge imposes and its
software consequences. Zhu et al. benchmark Hopper GPU-CC and identify
CPU-GPU transfer as the overhead source for LLM inference \citep{hopper-cc-benchmark};
Ibarra et al. add traffic and model-swap dynamics on H100 \citep{performance-cc-gpus};
Mohan et al. assess end-to-end CPU-GPU TEE readiness \citep{cpu-gpu-cc-ready}. We
extend this line three ways: a mechanism-level causal account (serialized
channels, revoked asynchrony, per-crossing tolls) rather than aggregate
overheads; the policy-inversion result with measured recoveries inside a
production serving runtime; and the Blackwell platforms, where the relative
bridge tax is larger and the multi-GPU fabric capability is new.

\textbf{Confidential serving systems.} PipeLLM overlaps speculative encryption
with GPU computation to hide bridge cipher costs on Hopper \citep{pipellm} -
complementary to our result, and an instructive contrast: PipeLLM adds
pipelining to the encryption path, while we show that \emph{existing} runtime
pipelining can need to be removed or restructured because the scarce resource
under CC is the serialized channel, not CPU cipher throughput (Section 4.3).

\textbf{LLM serving systems.} vLLM, SGLang, and DistServe optimize scheduling,
batching, KV management, and disaggregation for non-confidential GPUs [VLLM;
VLLM-PAGEDATTENTION; SGLANG-PAPER; DISTSERVE]. Our results are a case study
in how such policies interact with a changed hardware contract: the
optimization direction reverses when DMA concurrency disappears. Loader work
(safetensors, fastsafetensors \citep{safetensors,fastsafetensors}) similarly gains
a CC-specific dimension: secure context lifecycle becomes the dominant cost
to engineer around. Broader confidential-ML systematization appears in
\citep{ml-cc-sok}; platform trust building blocks in [INTEL-TDX; DMTF-SPDM;
PCI-TDISP].

\section{Limitations}

\textbf{Statistical scope.} B300 headline numbers are preserved single-run point
measurements from a time-bounded access window; the hardware has been
returned. We rest claims only on large effects (the smallest headline delta
we interpret is the 13\% serving row; mechanism claims rest on 2x-44x
effects), and we publish raw artifacts for every row in the repository evidence
index. Pro 6000 results come from longer campaigns with repeat and sweep
coverage.

\textbf{Mechanism attribution.} The roughly 330 microsecond per-crossing setup
cost is inferred from guest-side timing and profiler attribution, not from
driver or copy-engine tracing; the secure-channel serialization is observed
behaviorally (Section 4.2) and matches NVIDIA's documented context
restriction \citep{nvidia-secops}, but we cannot see channel scheduling directly.
CUPTI/Nsight-level instrumentation is the natural next step.

\textbf{Qualified recovery magnitude.} The B300 v10c worker-thread patch was
measured in one sweep campaign and its 92\% high-concurrency recovery remains
qualified; the one-flag recovery, which carries the paper's Blackwell claims,
was reproduced across three workload classes. An H200 port reproduced the
worker-thread mechanism (+6-7\%), but with a newer vLLM image and without the
same Blackwell inversion magnitude.

\textbf{Coverage.} n=4 fabric tenants are validated structurally but not under
CUDA workload; n=8 workload qualification was blocked by a host FM state bug;
NCCL collectives and CC-on vs CC-off NVLink overhead were not measured.
B300 serving used a newer vLLM image than Pro 6000 (SM100 requirement), so
cross-platform serving rows are context, not controlled comparisons.

\textbf{Generational scope.} The paper's primary measurements are Blackwell. The
H200 boundary experiment uses one physical Hopper GPU, paired CC-on/off
microbenchmarks, and a vLLM 0.22.1 serving check. It distinguishes mechanism
from Blackwell amplification, but it is not a full Hopper performance study.
It supports the bridge law and bounds the scheduling result: CC removes
async's benefit on Hopper, while Blackwell is where the loss becomes a true
policy inversion.

\section{Conclusion}

GPU confidential computing has been benchmarked as if its cost were a number.
It is better understood as a changed contract: compute stays native, but the
bridge between the confidential VM and the GPU becomes a serialized,
high-toll channel, and asynchrony across it is silently revoked. Software
built for the old contract - which is all software - converts that change
into large, workload-shaped losses: 26\% of dense-decode throughput, 131\% of
KV-restore latency, 34x of model-load time. Software built for the new
contract gets most of it back: a scheduling flag, a worker-thread drain, a
pooled-context loader, a reuse-aware offload policy. The bridge law that
predicts the losses also prescribes the fixes, and it held on both Blackwell
platforms we measured. An H200 boundary experiment shows the bridge law is not
a Blackwell artifact; the Blackwell result is that the same fixed bridge cost
becomes large enough, on faster serving hardware, to change which runtime
policy is correct.

Blackwell's second change is what the confidential unit is: B300 turns the
NVSwitch fabric into a tenant resource - 510 GB/s of GPU-to-GPU bandwidth
inside a CVM, partitioned and concurrent - while leaving its control plane
outside the tenant's evidence. The next confidential AI platforms will be
judged on both axes: whether their runtimes respect the serialized bridge,
and whether their fabrics become as attestable as their devices.

\section{Acknowledgments}

We thank Supermicro for providing access to the X14 systems evaluated in this
work through the Supermicro JumpStart Program.

\bibliographystyle{plainnat}
\bibliography{references}
\end{document}